\documentclass[useAMS,usenatbib]{mn2e}
\usepackage{bm}
\usepackage{graphicx}
\usepackage{epsf}

\def\apj{ApJ}                 % Astrophysical Journal
\def\aap{A\&A}                % Astronomy and Astrophysics
\def\mnras{MNRAS}             % Monthly Notices of the RAS
\def\nat{Nature}              % Nature
\def\solp{Solar~Physics} %Solar Physics

\usepackage{epsf}

\title[Relativistically expanding cylindrical electromagnetic fields]{Relativistically expanding cylindrical electromagnetic fields}
\author[K.N. Gourgouliatos ]{K.N. Gourgouliatos\thanks {E-mail: kng22@ast.cam.ac.uk}\\
Institute of Astronomy, University of Cambridge, Madingley Road, Cambridge CB3 0HA}

\begin{document}

\date{Accepted -. Received -; in original form -}

\pagerange{\pageref{firstpage}--\pageref{lastpage}} \pubyear{-}

\maketitle

%opening

\label{firstpage}
\begin{abstract}
We study relativistically expanding electromagnetic fields of cylindrical geometry. The fields emerge from the side surface of a cylinder and are invariant under translations parallel to the axis of the cylinder. The expansion velocity is in the radial direction and is parametrized by $v=R/(ct)$. We consider force-free magnetic fields by setting the total force the electromagnetic field exerts on the charges and the currents equal to zero. Analytical and semi-analytical separable solutions are found for the relativistic problem. In the non-relativistic limit the mathematical form of the equations is similar to equations that have already been studied in static systems of the same geometry. 
\end{abstract}


\section{Introduction}

The presence of magnetic fields in objects of astrophysical interest is evident. Starting from the Earth's magnetic field, to the solar magnetically driven phenomena and at the most extreme from magnetars to the recently observed filaments in the Perseus cluster of galaxies \citep{F2008}, magnetic fields play an important r\^ole in the evolution of the object they occur in. The study of the magnetic fields is complicated due to the form of partial differential equations that have to be solved and their non-linear behaviour. For this reason most of the research has been done on equilibrium configurations. Even in the non-relativistic limit a few of the published studies are analytical, for instance \cite{P1980, T1981, L1982, LB1994, A1994, VT1998} whereas most of them are computational i.e. \cite{WL1992, L2002, F2006} and references therein. In the relativistic case the studies are mainly motivated by pulsar magnetospheres, $\gamma$-ray bursts, and relativistic jets. Most of the solutions are numerical,  for instance \cite{LW1970, C1999, VK2003, G2004, K2006, M2006, UM2007, MN2007, KC2009} and only a few of them are analytic \cite{L1976,P2005, GL2008}.

In this paper we study the problem of a force-free relativistic electromagnetic field inspired by the studies previously made by \cite{A1994} for the geometry and the symmetry used; and from \cite{P2005} for the relativistic generalisation we make. \cite{P2005} showed that it is possible to study magnetic fields that expand uniformly within a sphere. These fields emerge from a point, obey the ideal MHD relation $\bm{E}=-\bm{v} \times \bm{B}$ and the net electromagnetic force is zero, then he solved the partial differential equation he found assuming a linear form. In this paper we show that an analogue case that permits analytical solutions and allows us to include time evolution exists for cylindrical magnetic fields. The field emerges from a linear singularity as in \cite{A1994} who studied this type of fields in the static limit, but in our problem the field expands radially. The expansion velocity is proportional to the distance from the axis and inversely proportional to time, thus every element of the plasma moves at constant speed. The expansion is slow near the axis and reaches the speed of light on a cylindrical surface. There is no field in the space outside this surface. The field viewed in the co-moving frame of reference obeys the force-free relation in the non-relativistic sense $(\nabla \times \bm{B}) \times \bm{B}=0$ as there is only magnetic field present, whereas in the frame of reference at rest with respect to the axis the net electromagnetic force by taking into account both the interaction of the magnetic fields with the currents and the interaction of the electric fields with the charges is zero. In the limit of low velocities the equation obtained is similar to the Grad-Shafranov partial differential equation studied for the case of magnetic fields in equilibrium in the absence of pressure, but for relativistic velocities they differ considerably. We proceed on the derivation of the equation and then we solve it for various cases.

\section{The problem}

Assume a non-resistive plasma of negligible inertia containing a magnetic field $\bm{B}$ and an electric field  $\bm{E}$ in cylindrical geometry $(R, \phi, z)$, these fields do not depend on $z$.  The electric field is related to the magnetic by the definition of ideal MHD $\bm{E}=-\bm{v} \times \bm{B}$, where $\bm{v}$ is the velocity of the plasma normalized to the speed of light. In addition assume that the plasma expands with
\begin{eqnarray}
\bm{v}=\frac{R}{ct}\bm{\hat{R}}\,.
\label{vel}
\end{eqnarray}
The magnetic field can be described by a flux function $P=P(v,\phi)$ for the $B_{R}$ and $B_{\phi}$ components, and leave without any constraint the $B_{z}$ component yet, apart from $\partial B_{z} /\partial z =0$
\begin{eqnarray}
\bm{B}=\nabla P(v,\phi) \times \bm{\hat{z}} +B_{z}(R,\phi,t) \bm{\hat{z}}\,,
\label{magnetic0}
\end{eqnarray}
or
\begin{eqnarray}
\bm{B}=\frac{1}{R}\frac{\partial P}{\partial \phi} \bm{\hat{R}}- \frac{1}{R} v \frac{\partial P}{\partial v} \bm{\hat{\phi}} +B_{z} \bm{\hat{z}}\,.
\label{magnetic}
\end{eqnarray}
The term arising from the gradient of $P$ will be called coplanar field $\bm{B}_{p}$. From the ideal MHD relation the electric field is
\begin{eqnarray}
\bm{E}=v\Big(B_{z} \bm{\hat{\phi}}+\frac{1}{R}v\frac{\partial P}{\partial v}\bm{\hat{z}}\Big)\,.
\label{electric}
\end{eqnarray}
The magnetic and the electric field must satisfy Maxwell's equations. Indeed, a magnetic field of the above form is by construction divergence-free. The induction equation is
\begin{eqnarray}
\nabla \times \bm{E}+\frac{1}{c}\frac{\partial \bm{B}}{\partial t}=0\,,
\label{induction0}
\end{eqnarray}
then by substituting equations (\ref{magnetic}) and (\ref{electric}) into equation (\ref{induction0}) we find that the equations for the $R$ and $\phi$ components are satisfied. In order the $z$ component of equation (\ref{induction0}) to be equal to zero, $B_{z}$ has to satisfy the following partial differential equation 
\begin{eqnarray}
2B_{z}+R\frac{\partial B_{z}}{\partial R}+t \frac{\partial B_{z}}{\partial t}=0\,,
\label{diffBz}
\end{eqnarray}
we re-express $B_{z}=g(t)W(v,\phi)$, substituting this into equation (\ref{diffBz}) gives
\begin{eqnarray}
t\frac{g'}{g}+2=0\,,
\end{eqnarray}
therefore $g\propto t^{-2}$, so $B_{z}$ can be written as $B_{z}=\frac{v^{2}}{R^{2}}W(v,\phi)$, where $W$ is an arbitrary function of $v$ and $\phi$. 

The two remaining Maxwell's equations are used to evaluate the charge ($\rho$) and the current density ($\bm{j}$), 
\begin{eqnarray}
\rho=\frac{\nabla \cdot \bm{E}}{4 \pi}\,,
\label{Gauss}
\end{eqnarray} 
\begin{eqnarray}
\bm{j}=\frac{1}{4 \pi}\Big(c\nabla \times \bm{B} -\frac{\partial \bm{E}}{\partial t}\Big)\,.
\label{induction}
\end{eqnarray}
As we are looking for fields which are force-free in this generalized sense we demand that the total force the electric and the magnetic fields exert to charge and current densities is zero. This force is
\begin{eqnarray}
\bm{F}=\rho \bm{E}+ \frac{\bm{j} \times \bm{B}}{c}\,.
\label{force}
\end{eqnarray} 
Thus, by substituting the current and charge density found above into equation (\ref{force}) we evaluate all three components of the force. From the $F_{z}$ component we take
\begin{eqnarray}
(1-v^{2})\frac{\partial W}{\partial \phi}\frac{\partial P}{\partial v}-\Big((1-v^{2})\frac{\partial W}{\partial v}-3vW\Big)\frac{\partial P}{\partial \phi}=0\,,
\label{zforce}
\end{eqnarray}
then we multiply equation (\ref{zforce}) by $(1-v^{2})^{1/2}$ and re-arragne the terms
\begin{eqnarray}
\frac{\partial([1-v^{2}]^{3/2}W)}{\partial \phi}\frac{\partial P}{\partial v} \nonumber \\
-\Big[(1-v^{2})^{3/2}\frac{\partial W}{\partial v}-3v(1-v^{2})^{1/2}W\Big]\frac{\partial P}{\partial \phi}=0\,.
\label{Jacobian1}
\end{eqnarray}
Setting $\beta=(1-v^{2})^{3/2}W$ equation (\ref{Jacobian1}) reduces to 
\begin{eqnarray}
\frac{\partial \beta}{\partial \phi}\frac{\partial P}{\partial v}-\frac{\partial \beta}{\partial v}\frac{\partial P}{\partial \phi}=0\,,
\label{Jacobian2}
\end{eqnarray} 
which is the Jacobian of $\beta$ and $P$ with respect to $v$ and $\phi$. As it equals zero, $\beta=\beta(P)$, thus $W=(1-v^{2})^{-3/2}\beta(P)$, where $\beta(P)$ is an arbitrary function of $P$. Given that, the $z$ component of the magnetic field becomes
\begin{eqnarray}
B_{z}=\frac{v^{2}}{R^{2} (1-v^{2})^{3/2}}\beta(P)\,.
\label{Bz}
\end{eqnarray}
Using equation (\ref{Bz}) in the force equation (\ref{force}) the $R$ and $\phi$ components of the force are both zero when 
\begin{eqnarray}
v^{2}(1-v^{2})\frac{\partial^{2}P}{\partial v^{2}}+v(1-2v^{2})\frac{\partial P}{\partial v}+\frac{\partial^{2} P}{\partial \phi^{2}}=-\frac{v^{2}}{(1-v^{2})^{2}}\beta \beta'\,.
\label{eqn}
\end{eqnarray}
This is the cylindrical version of Prendergast's equation, which is the basic equation deduced in this paper and our major aim hereafter is to find analytical and semi-analytical solutions. It simplifies if we apply the transformation $u=1/v$
\begin{eqnarray}
(u^{2}-1)\frac{\partial^{2}P}{\partial u^{2}}+u\frac{\partial P}{\partial u}+\frac{\partial^{2} P}{\partial \phi^{2}}=-\frac{u^{2}}{(u^{2}-1)^{2}}\beta \beta'\,.
\label{eqn1}
\end{eqnarray}
This is a convenient form of the force-free equation, which we shall use later in this paper to find analytical solutions. 

\section{Separable solutions}

In what follows we study equation (\ref{eqn}) and forms that allow separable solutions of the form 
\begin{eqnarray}
P=V(v)\Phi(\phi)\,.
\label{sol}
\end{eqnarray}
In order to do so, we investigate forms of $\beta$ which permit separation of variables. $\beta$ is any function of $P$ and this gives a great freedom on the solutions of the basic equation (\ref{eqn}), which may be highly non-linear if such a form is chosen. At first we study the form of the equation in the non-relativistic limit, then solutions in the absence of a $z$ component of the magnetic field and then solutions of the linearized form of equation (\ref{eqn}). 

\subsection{Non-relativistic limit}

For $v\ll 1$ equation (\ref{eqn}) reduces to 
\begin{eqnarray}
v^{2}\frac{\partial^{2}P}{\partial v^{2}}+v\frac{\partial P}{\partial v}+\frac{\partial^{2} P}{\partial \phi^{2}}=-v^{2}\beta \beta'\,,
\label{nonrel}
\end{eqnarray}
Separable solutions are allowed in two cases. In the linear case for $\beta \beta' =c_{0}^{2}P$, equation (\ref{nonrel}) becomes
\begin{eqnarray}
\frac{v}{V}\frac{d}{dv}\Big(v\frac{d V}{dv}\Big)+c_{0}^{2}v^{2}=-\frac{1}{\Phi}\frac{d^{2}\Phi''}{d\phi^{2}}=c_{1}^{2}\,,
\label{lnonrel}
\end{eqnarray}
where $c_{1}$ is the constant of separation and it is a real number as $\Phi$ has to be periodic. The solution to the angular part is 
\begin{eqnarray}
\Phi=c_{A}\cos(c_{1}\phi)+c_{B}\sin(c_{1}\phi)\,.
\end{eqnarray}
The differential equation for $V$ admits Bessel functions of order $c_{1}$ for solutions
\begin{eqnarray}
V(v)=c_{C}J_{c_{1}}(c_{0}v)+c_{D}Y_{c_{1}}(c_{0}v)\,.
\end{eqnarray}
The constants $c_{A}$, $c_{B}$, $c_{C}$ and $c_{D}$ can be evaluated subject to the boundary conditions. The horizontal shear causing the $z$ component of the field is parametrized by the constant $c_{0}$.

The other class of separable solutions is for $V=V_{0}v^{-l}$ and $\beta \beta' =c_{0}^{2} P^{1+2/l}$derived by \cite{A1994}, who studied a problem of the same geometry in the static case. Indeed if we substitute the forms above in equation (\ref{nonrel}), the terms involving $v$ cancel and we take an ordinary differential equation for $\Phi$
\begin{eqnarray}
\Phi''+l^2\Phi+c_{0}^{2}V_{0}^{2/l}\Phi^{1+2/l}=0\,.
\label{fnonrel}
\end{eqnarray}
This equation can be solved numerically, subject to the boundary conditions and the parameters $c_{0}$, $l$ and $V_{0}$. It is possible to by-pass the numerical solution of equation (\ref{fnonrel}) and construct an approximate solution for $l$ close to zero as the one found by \cite{LB1994}, by normalizing to unity $f=\Phi/\Phi_{\rm{max}}$ we find
\begin{eqnarray}
f=1-\frac{l+1}{l}\ln\cosh(d(\phi-\phi_{\rm{max}}))
\end{eqnarray}
where $d^{2}=\frac{l+1}{l}c_{0}^{2}(V_{0}\Phi_{\rm{max}})^{2/l}$ and $\phi_{\rm{max}}=\frac{1}{d}\cosh^{-1}\exp\frac{l}{l+1}$. If $l$ is no more small then we can use the more accurate and compicated form described in appendix A of \cite{L2006}.

Although the mathematical treatment described above is the same to Aly's case, the two equations describe distinct physical configurations. The fields in our problem are not static but expand while being force-free. This growth can also be conceived as some flux generated from the base and expanding with velocity equal to $R/{ct}$ . In Aly's case the fields are static, however it is possible to achieve expansion due to the shearing of the field lines on the surface of the cylinder. That problem is studied as a sequence of static solutions where the transition from one solution to the next one takes place slowly enough so that the displacement current is negligible. In our case the shearing does not change explicitly with time as it is imposed from the beginning. However it is possible to assume the field lines to be sheared slowly, indeed as discussed by \cite{L2006} MHD systems where the boundary conditions change slowly have negligible induction currents and pass through a sequence of equilibrium stages. In our case, if the shearing of the field lines takes place slowly, then a series of force-free solutions of expanding fields describes this process. Thus the shearing of the field lines may take place simultaneously with the expansion.

\subsection{Relativistic Solutions}

Here we investigate solutions of the fully relativistic equation (\ref{eqn}) without neglecting any term. First we study solutions in the absence of a $z$ component for the magnetic field which gives a coplanar magnetic field without any current and charge density and then solutions containing all three components by assuming a linear form for the right hand side of the differential equation. 

\subsubsection{Current-free solution}

The simplest non-trivial solution of equation (\ref{eqn}) exists in the absence of a $z$ component of the magnetic field, $\beta=0$. It is a magnetic field that emerges from the surface of a cylinder where no shear is imposed. In this configuration there is no current or electric charge. We shall use equation (\ref{eqn1}) as this form is more convenient, let $P=V(u)\Phi(\phi)$ and derivatives to be denoted by dash, equation (\ref{eqn1}) becomes
\begin{eqnarray}
\frac{(u^{2}-1)V''+uV'}{V}=-\frac{\Phi''}{\Phi}=c_{1}^{2}\,.
\label{currentfree}
\end{eqnarray}
The solution of the angular part is of sinusoidal form 
\begin{eqnarray}
\Phi=c_{A}\sin(c_{1}\phi)+c_{B}\cos(c_{1}\phi)\,,
\end{eqnarray}
the solution for $V(u)$ is 
\begin{eqnarray}
V=c_{C}(u+\sqrt{u^{2}-1})^{c_{1}}+c_{D}(u-\sqrt{u^{2}-1})^{c_{1}}\,.
\end{eqnarray}
This corresponds to a coplanar magnetic field, where every magnetic field line lies on a plane normal to the axis of the cylinder.

\subsubsection{Linear solution}

Unlike the non-relativistic equation, which admits two classes of separable solutions when $\beta \neq 0$, in the relativistic regime, separable solutions exist only in the linear case, and there is no analogue to the non-relativistic powerlaws. To show this assume separable solutions of the standard form; in division of equation (\ref{eqn1}) by $P$ it is 
\begin{eqnarray}
\frac{(u^{2}-1)V''+uV'}{V}+\frac{\Phi''}{\Phi}=-\frac{u^{2}}{(u^{2}-1)^{2}}\frac{\beta \beta'}{P}\,,
\label{divided}
\end{eqnarray}
the first term of the left hand side is only a function of $u$ and can be expressed as $F(V)$, the second term similarly is $G(\Phi)$, as for the right hand side let $H(P)=\beta\beta'/P$. The next step is to act with the operators $\Phi d/d\Phi$ and in the resulting expression with the operator $V d/dV$; the final equation is
\begin{eqnarray}
V\frac{d (\frac{u}{u^{2}-1})^{2}}{dV}\frac{dH}{d \ln P}=-\Big(\frac{u}{u^{2}-1}\Big)^{2}\frac{d^{2}H}{d (\ln P)^{2}}\,.
\end{eqnarray}
Then name $Y=\frac{u^{2}}{(u^{2}-1)^{2}}$ and substitute above,
\begin{eqnarray}
\frac{V}{Y}\frac{dY}{dV}=-\frac{d^{2}H}{d (\ln P)^{2}}/\frac{d H}{d \ln P}=n\,.
\label{separation}
\end{eqnarray}
From equation (\ref{separation}) it is $Y\propto V^{n}$ which leads to $V=V_{0}(\frac{u}{u^{2}-1})^{2/n}$ and the form of $H(P)$ that permits separation of variables is $H=c_{1}+c_{2}P^{-n}$. When substituting these forms in the right hand side of equation (\ref{eqn1}) it becomes a sum of a term that only depends on $u$ and one that only depends on $\phi$, however when the left hand side of the equation is considered the terms involving $u$ do not sum up to a constant thus it is impossible to separate the equation by using power laws. It is only the linear form for $\beta \beta'=c_{0}^{2}P$, which permits separation of variables, indeed
\begin{eqnarray}
\frac{(u^{2}-1)V''+uV'}{V}+c_{0}^{2}\frac{u^{2}}{(u^{2}-1)^{2}}=-\frac{\Phi''}{\Phi}=c_{1}^{2}\,.
\label{linear}
\end{eqnarray}
The angular part admits again sinusoidal solutions, as for the $V(u)$ the equation to solve is
\begin{eqnarray}
\frac{(u^{2}-1)V''+uV'}{V}+c_{0}^{2}\frac{u^{2}}{(u^{2}-1)^{2}}=c_{1}^{2}\,,
\label{linearV}
\end{eqnarray}
which cannot be integrated analytically and we solve it numerically.

The numerical solution of equation (\ref{linearV}) depends on the choice of the parameters and the boundary conditions. Assume that some flux emerges from a cylindrical surface $R_{0}$ at time $t_{0}$ which in velocity space lies $v_{0}=R_{0}/(ct_{0})$, the boundary conditions for $V$ on this surface are determined by the fields. The fields are expressed in terms of $V$ and $\Phi$, they are 
\begin{eqnarray}
B_{R}=\frac{1}{R}V\Phi'\,,
\label{BR}
\end{eqnarray}
\begin{eqnarray}
B_{\phi}=-\frac{v}{R}V'\Phi\,,
\label{Bf}
\end{eqnarray}
\begin{eqnarray}
B_{z}=\frac{c_{0}v^{2}}{R^{2}(1-v^{2})^{3/2}}V\Phi\,.
\label{B_z}
\end{eqnarray}
The value of $V$ on the surface the flux emerges from is related to the intensity of the $R$ component of the magnetic field and the derivative $V'(v_{0})$ to the intensity of $\phi$ component, in what follows we use the normalised value $V(v_{0})=1$. Then the parameters $c_{0}$ and $c_{1}$ are chosen and we integrate the differential equation. The physical meaning of the parameter $c_{0}$ is related to the horizontal shear of the field lines and therefore since $B_{z}$ is proportional to $c_{0}$, the more shear is induced on the base of the field the larger this parameter is. An inspection of the differential equation even without proceeding to the numerical solution demonstrates the decelerating effect of this term, greater shear leads to a stronger $B_{z}$ component of the magnetic field. This component is proportional to $\gamma^{3/2}$ and exerts a strong force to the magnetic field. Thus, the flux decreases rapidly for $v$ close to unity. After this point the solution undergoes oscillations (Figures 1, 2) and the field forms disconnected appendages. These structures are the result of the linear assumption made previously. It is now possible to draw the field lines for any time $t_{1}$ (Figures 3, 4) as we have the expressions for $V$ and $\Phi$ and the field lines are given by substituting expressions (\ref{BR}), (\ref{Bf}) and (\ref{B_z}) in
\begin{eqnarray}
\frac{dR}{B_{R}}=\frac{R d \phi}{B_{\phi}}=\frac{dz}{B_{z}}\,.
\label{fieldlines}
\end{eqnarray}
\begin{figure}
  \centering
 \includegraphics[width=0.5\textwidth]{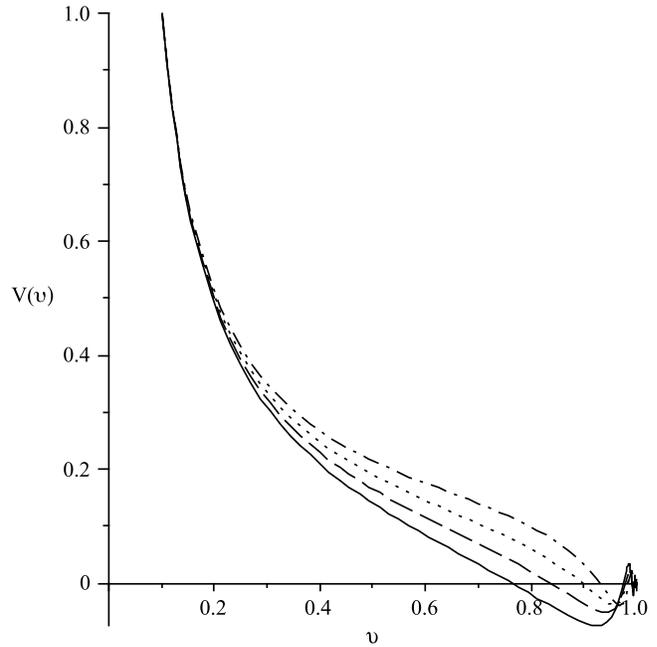}
     \caption{The numerical solutions of equation (\ref{linear}). $V$ is normalized to unity at $v_{0}=0.1$, $c_{0}=1$ and $c_{1}=1$, the solid line is for $V'(v_{0})=-10.1$, the dashed for $V'(v_{0})=-10$, the dotted for $V'(v_{0})=-9.9$ and the dash-dotted for $V'(v_{0})=-9.8$. This decrease in the absolute value of the derivative leads to a weaker $B_{\phi}$ at $v_{0}$, thus the field emerges closer to the perpendicular and the flux function reaches a greater $v_{\rm{max}}$ before starting the oscillations.}
\end{figure}
\begin{figure}
  \centering
 \includegraphics[width=0.5\textwidth]{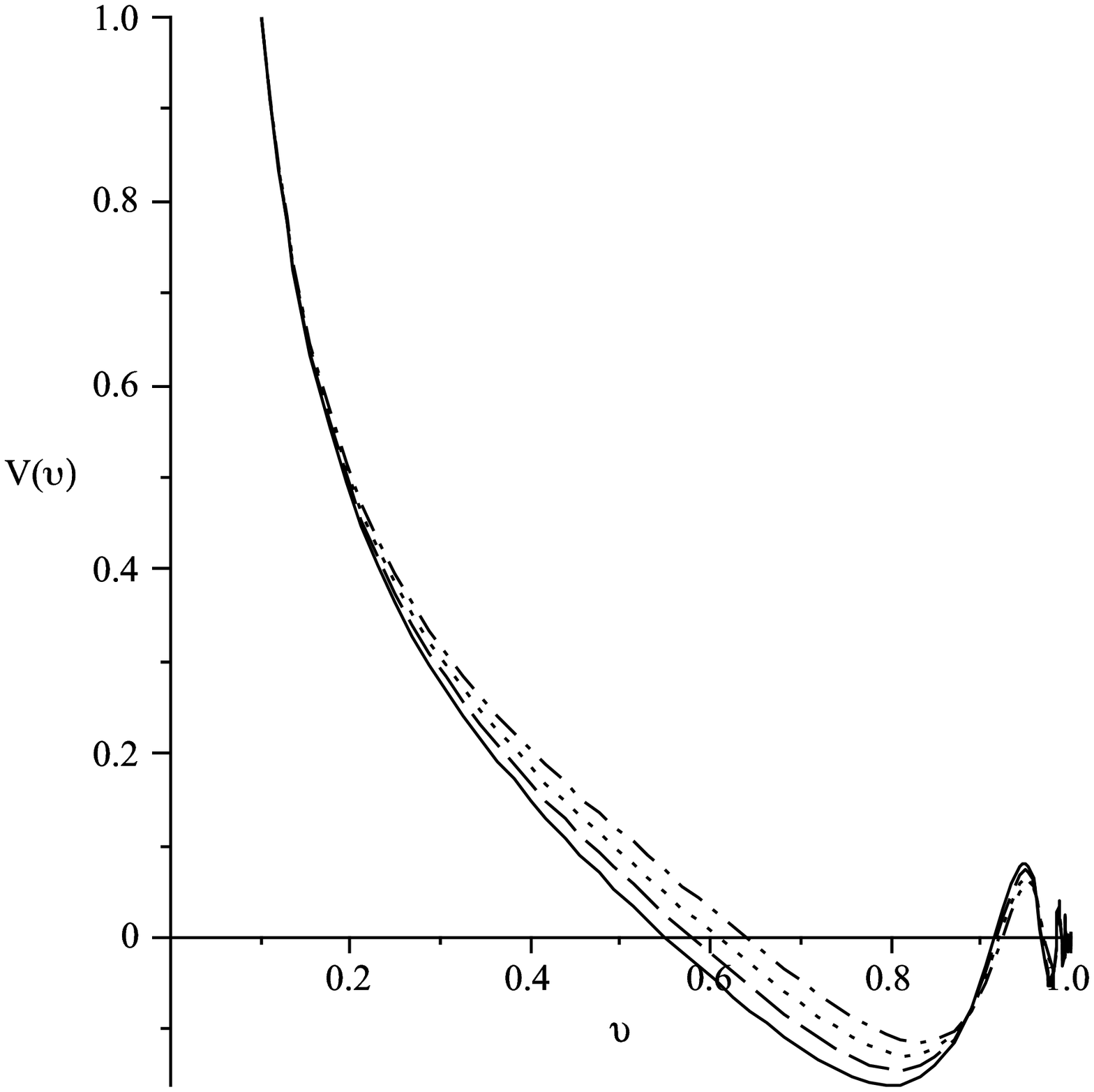}
     \caption{The numerical solutions of equation (\ref{linear}). $V$ is normalized to unity at $v_{0}=0.1$, $c_{0}=2$ and $c_{1}=1$, the solid line is for $V'(v_{0})=-10.1$, the dashed for $V'(v_{0})=-10$, the dotted for $V'(v_{0})=-9.9$ and the dash-dotted for $V'(v_{0})=-9.8$. This decrease in the absolute value of the derivative leads to a weaker $B_{\phi}$ at $v_{0}$, thus the field emerges closer to the perpendicular and the flux function reaches a greater $v_{\rm{max}}$ before starting the oscillations. Compared to the lines plotted in Figure 1 the $z$ field is stronger here and the oscillations start earlier.}
\end{figure}
\begin{figure}
  \centering
 \includegraphics[width=0.5\textwidth]{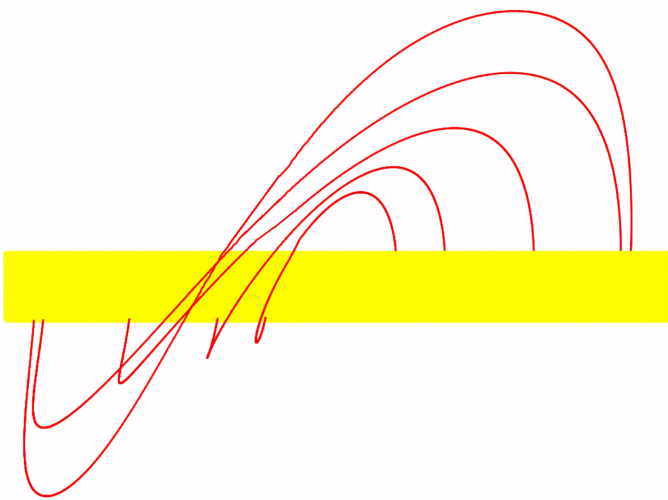}
     \caption{The magnetic field lines. They emerge from the yellow cylindrical conductor and they expand radially. The ones reaching greater distance from the cylinder have more shear.}
\end{figure}
\begin{figure}
  \centering
 \includegraphics[width=0.5\textwidth]{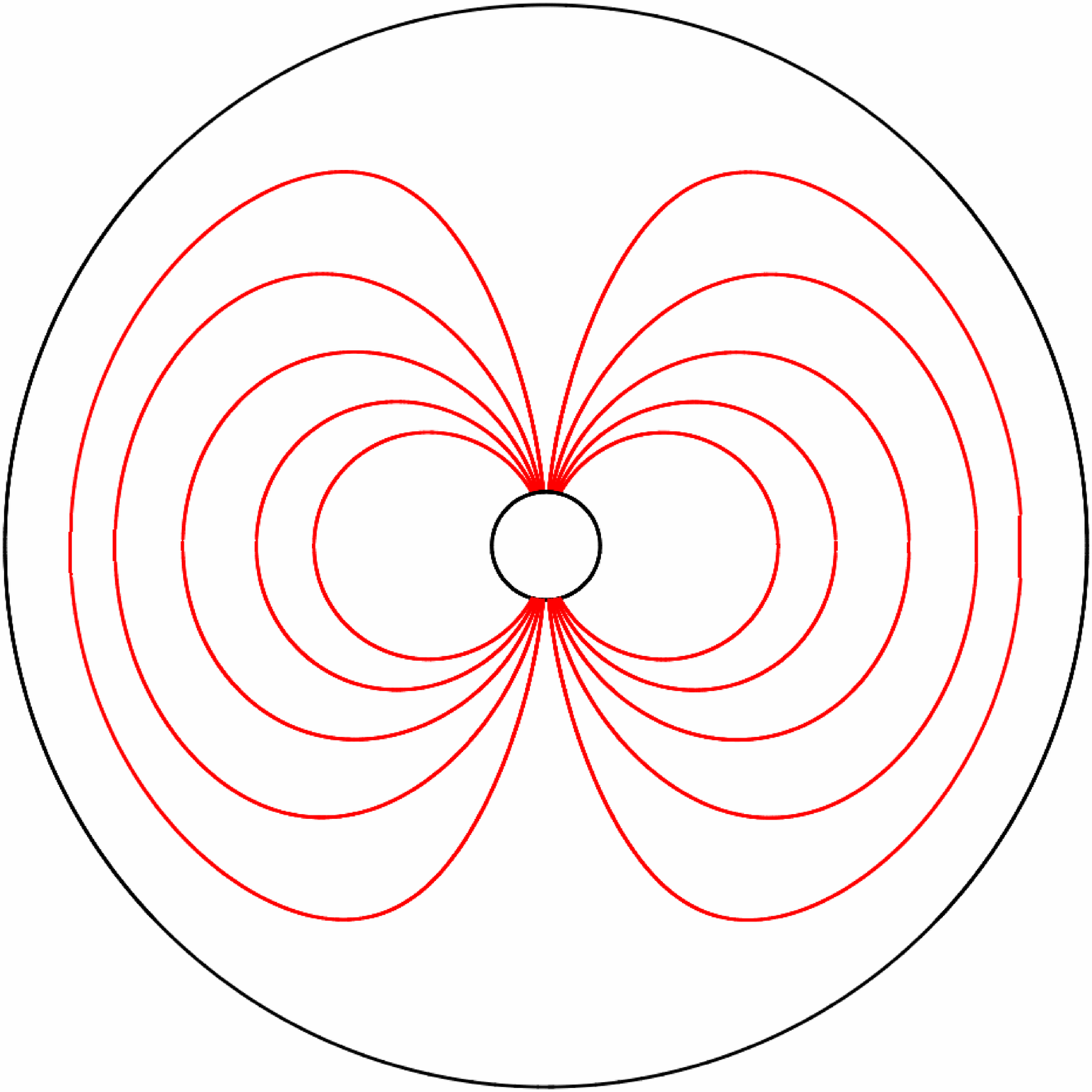}
     \caption{Surfaces of constant $P$, the line of sight is parallel to the axis of the cylinder. The inner circle corresponds to a cylinder of radius $v=0.1$ and the outer circle to one of radius $v=1$.}
\end{figure}
The other parameter $c_{1}$ defines the order of the multipole chosen, for instance $c_{1}=1$ is a cylindrical dipole and $c_{1}=2$ gives a quadrupole etc. If a non-integer value is chosen, then the problem can be studied within a confined arcade, of any opening angle, however the solution will not have a physical significance outside this domain. Finally the value of the derivative $V'(v_{0})$ is chosen and the differential equation is integrated numerically. In the solutions plotted we choose four values for it, we find that for a smaller absolute value of $V'(v_{0})$ the flux reaches greater distances in $v$ space; thus fields where the ratio $B_{R}(v_{0})/B_{\phi}(v_{0})$ is bigger have fluxes that reach greater distances, so the more radial the flux is at the base the further it gets before it starts oscillating. 

\section{Physical qunatities}

In this section we study the shear and the energy stored in this configuration subject to $c_{0}$. We assume that the same radial flux emerges from the surface by keeping $V(v_{0})$ the same and it reaches a maximum velocity $v_{\rm{max}}$ also constant, where $V(v_{\rm{max}})=0$, given these boundary conditions we solve equation (\ref{linearV}) for various values of the parameter $c_{0}$. Then we use the solutions to study the shear and the energy.

\subsection{Shear}

The amount of shear for a given field line can be evaluated by integrating equation (\ref{fieldlines}) after substituting the fields from equations (\ref{BR}-\ref{B_z}). An element of a field line labeled by $P_{i}$ that lies between $v$ and $v+dv$ in velocity space is sheared by
\begin{eqnarray}
dz=-\frac{c_{0}P_{i}vdv}{V(1-(P_{i}/V)^{2})^{1/2}(1-v^{2})^{3/2}}\,.
\label{shear}
\end{eqnarray}
An important feature of this quantity is that it does not change as the field expands. We evaluate the total shear of a field line by integrating equation (\ref{shear}) from $v_{0}$ where the field line emerges, up to the maximum $v_{t}$ it reaches, this $v_{t}$ is given by solving equation $V(v_{t})=P_{i}$, then by symmetry the shear from the top to the other foot point is the same. Inspection of equation (\ref{shear}) shows that the shear depends on $c_{0}$ as it appears directly and indirectly in the integral through the form of $V$ which is a numerical solution depending on the parameter $c_{0}$, there is also a dependence on the choice of the field line which also appears directly in the equation but also indirectly through the end point of integration $v_{t}$ which depends on $P_{i}$. In Table 1 we use the results of the linear solution to evaluate the shear $(Z)$ of the field line which has been stretched the most for a given $c_{0}$. The numerical integration shows actually that as $c_{0}$ increases the shear increases, a proportionality relation between $c_{0}$ and $Z$ is a first approximation, and the deviation of this approximation is due to the indirect dependence on $c_{0}$ through the other quantities appearing in (\ref{shear}).

\subsection{Energy}

\subsubsection{Energy flow}

The total energy carried by an electromagnetic field in a volume $V_{i}$ is
\begin{eqnarray}
E_{tot}=\frac{1}{8 \pi}\int (\bm{B}^{2}+\bm{E}^{2}) dV_{i}\,.
\label{energy0}
\end{eqnarray}
We can express the forms of the electric and the magnetic field in terms only of $t$ and $v$ and study the energy contained in a volume in velocity space and how it evolves with time. By substituting the forms of the magnetic fields from equations (\ref{BR}-\ref{B_z}) and by the fact that $\bm{E}=-\bm{v} \times \bm{B}$ the energy contained in a volume in velocity space with boundaries $v_{0}$ and $v_{\rm{max}}$, $z$ from $0$ to $z_{1}$ and $\phi$ from $0$ to $2\pi$, is
\begin{eqnarray}
E_{tot}=\frac{z_{1}}{8 \pi}\int^{2 \pi}_{0}\int_{v_{0}}^{v_{\rm{max}}}\Big\{ \frac{V^{2} \Phi'^{2}}{v^{2}}\nonumber \\
+(v^{2}+1)\Big[V'^{2}\Phi^{2}+\frac{c_{0}^{2}V^{2}\Phi^{2}}{(ct)^{4}(1-v^{2})^{3}}\Big]\Big\}vdvd\phi\,.
\label{energy}
\end{eqnarray}
The form of the integral in equation (\ref{energy}) suggests that the energy within a constant volume in velocity space changes with time. It is constant only when there is no $z$ component of the magnetic field. The Poynting vector of an electromagnetic field is
\begin{eqnarray}
\bm{S}=\frac{c}{4\pi}\bm{E} \times \bm{B}\,,
\end{eqnarray}
and by substituting in terms of $V$ and $\Phi$ we find
\begin{eqnarray}
\bm{S}=\frac{cv}{4\pi}\Big\{\Big[\frac{V'^{2}\Phi^{2}}{(ct)^{2}}+\frac{c_{0}^{2}V^{2}\Phi^{2}}{(ct)^{4}(1-v^{2})^{3}}\Big]\bm{\hat{R}}\nonumber \\
+\frac{VV'\Phi\Phi'}{v(ct)^{2}}\bm{\hat{\phi}}-\frac{c_{0}V^{2}\Phi\Phi'}{v(ct)^{3}(1-v^{2})^{3/2}}\bm{\hat{z}}\Big\}\,.
\label{Poynting}
\end{eqnarray}
As the $R$ component of the Poynting vector is positive, the energy flows outwards as the fields expand.

\subsubsection{Energy as a function of the shear}

We have integrated equation (\ref{energy}) from $v=0.1$ to $v_{\rm{max}}=0.95$ for a range of values of $c_{0}$ at $ct=1$. We used the form of $V$ found by solving the differential equation (\ref{linearV}) subject to the boundary conditions stated in this section; the results appear in Table 1. The overall conclusion is that the energy increases with $c_{0}$. This is due to two reasons, the $z$ component of the magnetic field increases, thus it carries more energy. This increase on $B_{z}$ has also a side effect, for the field to remain force-free the coplanar component of the magnetic field $\bm{B}_{p}$ and $\bm{E}_{z}=-\bm{v} \times \bm{B}_{p}$ have to increase to balance the extra force. Thus the energy carried by them becomes larger as well. It is evident from Table 1 that both $En_{p}$ and $En_{z}$ increase with $c_{0}$ and so does their sum $E_{tot}$.

\begin{table}
 \centering
 \begin{tabular}{|r|r|r|r|r|}
\hline
$c_{0}$ & $Z$ & $En_{p}$   & $En_{z}$ & $E_{tot}$ \\
\hline
0.0      & 0.000      & 1.000  & 0.000  & 1.000 \\
0.1      & 0.175      & 1.000  & 0.001  & 1.001 \\
0.2      & 0.354      & 1.001  & 0.002  & 1.003 \\
0.3      & 0.532      & 1.003  & 0.004  & 1.007 \\
0.4      & 0.738     & 1.006   & 0.008  & 1.014 \\
0.5      & 0.934     & 1.011   & 0.014  & 1.025 \\
0.6      & 1.160     & 1.019   & 0.024  & 1.043 \\
0.7      & 1.426     & 1.034   & 0.039  & 1.073 \\
0.8      & 1.690     & 1.064   & 0.068  & 1.132 \\
0.9      & 2.057     & 1.131   & 0.131  & 1.262 \\
1.0      & 2.332     & 1.336   & 0.312  & 1.648 \\
\hline
\end{tabular}
\caption{The physical quantities of the system for $c_{0}$ ranging from $0$ to $1$, evaluated by using the solution of equation (\ref{linearV}) subject to the boundary conditions $V(0.1)=1$, $V(0.95)=0$ and $c_{1}=1$. As $c_{0}$ increases the shearing of the field lines $(Z)$ increases. The energy carried by the field also increases, where $En_{p}$ is the energy carried by the coPlanar component of the magnetic field and the $E_{z}$ of the electric; $En_{z}$ is the energy carried by $B_{z}$ and $E_{\phi}$. }
 \label{Table 1}
\end{table}

\section{Applications}

Observations of $\gamma$-ray flares \citep{Pal2005} suggest that they are associated to strong magnetic fields expanding with relativistic velocities. These flares are thought as potential origins of short duration $\gamma$-ray bursts \citep{H2005}. The relativistic solution of the equations studied in this paper can be applied to the initial stages of magnetar giant flares emerging. It has been proposed \citep{Lyu2006} that giant flares from $\gamma$-ray repeaters are formed through processes similar to coronal mass ejections on the Sun. As opposed to solar arcades, they reach high Lorentz factors. Our model describes the electromagnetic field of arcades which expand and reach relativistic velocities. When the flare expands with a great velocity it is essential to take into account the relativistic effects as extra terms appear on equation (\ref{eqn}) compared to the non-relativistic form, equation (\ref{nonrel}). In our study the fields expand only in one dimension, this is a reasonable assumption when the magnetic arcades are not large compared to the radius of the object where they emerge from or when there is a primary direction of expansion. An other issue is the neglect of inertia forces, this is fine when the plasma is underdense and most energy is carried by the magnetic field. However very close to the speed of light, no matter how small the density is, it will be no more negligible as it will be multiplied by a large Lorentz factor. When the expansion in a second dimension becomes important then the properties of the spherical geometry have to be taken into account as they are presented by \cite{P2005, GL2008}. 

\section{Conclusions}

In this paper we derived the force-free analogue of Prendergast's equation \citep{P2005} in the case of cylidrically expanding magnetic fields. These solutions offer a theoretical insight to both relativistic and non-relativistic expanding systems. The non-relativistic solutions have structural similarities to static ones where time is not explicitly included, but it is implied via a series of static solutions where the expansion is merely due to shearing of the magnetic field lines whereas the flux emerging from the surface is constant. In our case the field expands due to two reasons: shearing of the field lines as in the static problem and due to an increase of the flux emerging from the surface. The increase of flux appears directly in the equations.

The plane parallel geometry also sets constraints to the applications of these solutions. Astrophysical systems are more likely to occur in spherical geometry, a case studied previously \citep{GL2008}, however this particular geometry leads to results of satisfactory accuracy compared to those of the spherical geometry when the radius the magnetic flux extends is comparable to the radius of curvature of the surface the magnetic field emerges. Systems where the expansion is mainly in one dimension are also described by this geometry. An astrophysical system of interest is that of solar arcades, this model describes the combined effect of the extra flux emerging from the base and of the shearing of the field lines provided that the size of the arcade is small compared to the radius of the Sun. A detailed comparison of plane parallel models and spherical ones can be found in \cite{G2008} and the loss of equilibrium in these models leading to the opening of the field lines has been studied by \cite{U2002}. It is also shown that the energy contained in the magnetic arcade increases when more shear is imposed. 

In relativistic systems the demand of simultaneous expansion and shearing has some problems as the field at the top expands very fast whereas the shearing of the field lines at the base has to take place slowly. Thus the message for the $B_{z}$ component to increase will arrive much later at the top. For this reason we consider that the magnetic field is sheared before the expansion takes place. 
 
\section*{Acknowledgements}

The author is grateful to Professor Donald Lynden-Bell for the inspiring discussions and guidance. 

{}

\bsp

\label{lastpage}

\end{document}